# Transferring Voice using SMS over GSM Network

M. Fahad Khan, Saira Beg

**Abstract**—The paper presents a methodology of transmitting voice in SMS (Short Message Service) over GSM network. Usually SMS contents are text based and limited to 140 bytes. It supports national and international roaming, but also supported by other telecommunication such as TDMA (Time Division Multiple Access), CDMA (Code Division Multiple Access) as well. It can sent/receive simultaneously with other services. Such features make it favorable for this methodology. For this an application is developed using J2ME platform which is supported by all mobile phones in the world. This algorithm's test is conducted on N95 having Symbian Operating System (OS).

**Index Terms**—GSM (Global System for Mobile Communications) Network, Voice, SMS (Short Message Service).

## 1 Introduction

Mobile technology is, perhaps, the most rapidly growing technology. In 2G network, GSM (Global System for Mobile Communications) have more significant contribution than other digital technologies [1-2]. The services provided by the GSM are grouped into three main categories such as: Tele services (TS), Bearer services (BS) and Supplementary services (SS). TS service concerns with regular telephone calls, emergency calls and voice mails etc. BS services are all about data services such as SMS (Short Message Service), cell broadcast etc. SS are value added features; calls cost, call hold, call waiting etc [3].

SMS and MMS (Multimedia Messaging Service) are used for delivering short messages in GSM. Almost all cell phones (GSM devices) are capable of sending and receiving SMS, following stored-and-forward way of mechanism. It is much cheaper than dialing. Each message can not be more than 160 characters and characters should be alphanumeric or binary non-text. SMS have international and national roaming and it is supported by all major technologies. Although MMS contain richer contents than SMS and have no size limit, but it needs 3G network capability for sending large size messages [3-4].

In this paper we present a method of transferring voice using SMS. We developed an application which enable voice message on all GSM devices, even those which does not have GPRS (General Packet Radio Service) or other 3G capability. In this framework, our main focus is to reduce hardware dependencies and provide an alternative way of transferring one's voice to the receipts. This can reduce language barrier and save time of keying or typing SMS. This Paper is divided into following sections. Section 2, 3 represents SMS utilization in other areas and related work respectively, following this section 4 represents proposed methodology, results discussion and comparison, at the last we conclude the paper.

## 2 SMS Utilization In other Areas

SMS is used widely over the globe. It supports national and international roaming, low rates (cost per SMS) and simultaneously sent/ received with other services over GSM [5] make it useable for business and in mobile banking etc. [6] represents the SMS banking. With this method user can bank with their phones 24 hours of a day. Authors developed banking agent, which processed the requests coming from the clients and send responses to them. [7] describes SMS based Home Appliance Control System (HACS) for automating appliances and security. Author shows a novel way of monitoring and controlling appliances through SMS. This system is based on three main modules; PC contains HACS software which controlled the home appliances, GSM Modem gives capability of sending/receiving SMS to and from the system and third component is mobile device which communicates with GSM modem. Quantitative research regarding SMS marketing is presented in [8]. Research shows that discounts and promotional offer based SMS have positive response.( But simple used by them is too small which does not give the clear picture yet). [9] gives the idea of using SMS based appointment remainder to outdoor patients. Results suggested that with this attendance of outpatients is improved and it also brings the satisfaction among patients regarding that hospital.

[5] presents MMS based system for traffic police. In this system, photo of high speedy vehicle is sent to the centralized police data base in a real time by using MMS service of GSM. Recently, GSM network is used for health care system. Architecture of that system is based on three layers; top layer is used for web services, bottom layer is for

- *M.Fahad.Khan. is with the Federal Urdu University of Arts, Science and Technology, Islamabad. E-mail: mfahad.bs@gmail.com.*
- *Saira.Beg. is with the COMSATS Insitiute of Information Technology, Islamabad. E-mail: sairabegbs@gmail.com.*

healthcare and middle layer control the messages flow between top and bottom layer [10].

## 3 RELETED WORK

Different research articles are already being presented regarding this idea. [11-12] suggests the transfer of voice through SMS. In this author has used the utterance generated by the encoder card present in a mobile phone. After getting utterance, he converts it into a non text representation, insert that text into the body of SMS and send. At receiver end, firstly extract the text than synthesize audio representation from it and lastly play that audio through audio device (speaker) of a mobile phone. [13] represents transfer and display animation through SMS over a mobile network. Creating such SMS, it needs mobile identification number by user in order to identify receiving party, an animation flag (yes/no), if yes than specify content pointer location for animation at receiver side and animation type. Lastly, user enters the actual text of the SMS. For displaying, receiving device parses the SMS first in order to determine desire animation characteristics, animation flag is checked, content location is determined from location content pointer. Now merge SMS text, content and apply animation type specified by the sender. [14] discuss the two ways of sending SMS intended for access terminal. If data session exists between the device and access terminal, embed that SMS into data message, and transmit over data network. In second case if data session is not present than send that SMS via voice network.

## 4 PROPOSED METHODOLOGY

Our application running on a mobile takes voice input from the user. The user only has to press the 'send' button just like in case of sending an SMS. The application hides remaining steps between recording a voice input and converting it into an SMS. The application first captures and stores the user input in a ByteArrayOutputStream. It then converts that signed ByteArrayOutputStream into unsigned integer array. The Extended ASCII character values ranging from 0 to 31 cannot be sent through SMS because they are reserved for special functions like 'null' etc. So, our application adds 256 to the unsigned Integer Array values falling in this range to move them on to the range 256 to 287. The application applies this approach to convert the array values into their respective Extended ASCII values. Finally, convert these characters into strings and set message indexes. It, then, sets these strings as payload text of SMS and sends to receiver. When this message is received on the other side, where reverse procedure is applied on the received data to convert it back into a voice message. Figure 1 represents the block diagram of the methodology.

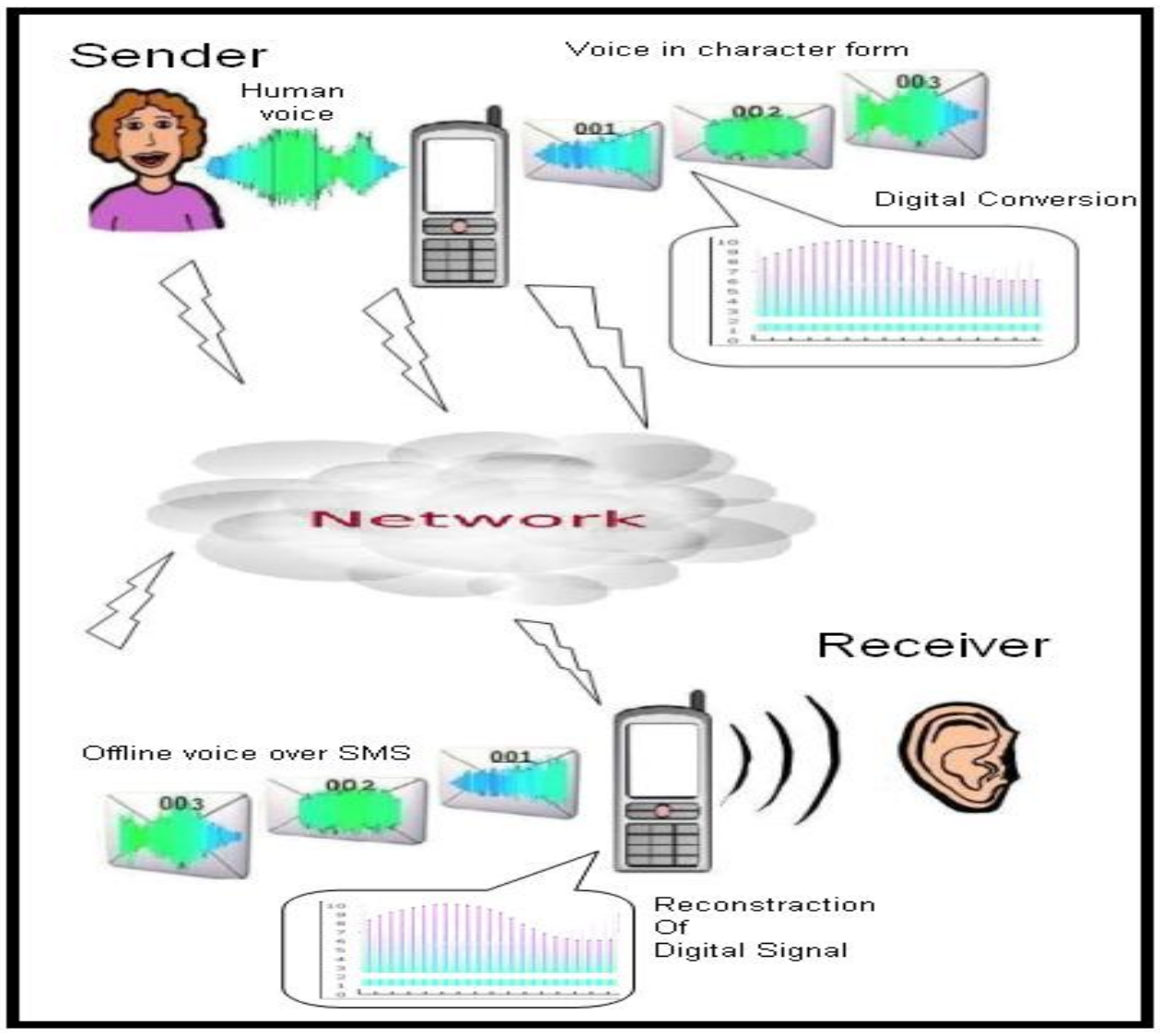

Fig. 1. Architecture of Proposed Methdology

### 4.1 Results and Discussion

We developed application using j2me platform and used N95 mobile phone which having Symbian Operating System (OS) for results. We consider four main factors; number of words spoken by user, number of characters against those words, number of messages required for compensating those characters and number of connected messages. As SMS have 140 bytes approximately which can transfer over GSM network. Using concatenation of SMS [15] procedure we reserve first three characters for indexing. This gives 000-999 connected SMS indexing. Indexing is used for placing received SMS in order at receiver end. We used three formats; PCM (Pulse-Code Modulation), ULAW (Companding algorithm) and AMR (Adaptive Multi-Rate). PCM is standard form for digital audio in computers. The results shows that ARM format which is audio data compression scheme has less number of characters, messages and connected messages while PCM and ULAW have approximately same. Table1 shows different parameters on which results are conducted. We test this application on different human voices in order to judge variation of the results. As each voice has its own texture which can produce variation.

TABLE 1 REPRESENTING MAIN PARAMETERS

| *Test #* | *Words #* | *Number of Characters* | | | *Number of Messages* | | | *Number of Connected Messages* | | |
|---|---|---|---|---|---|---|---|---|---|---|
| | | *PCM* | *ULAW* | *AMR* | *PCM* | *ULAW* | *AMR* | *PCM* | *ULAW* | *AMR* |
| 1 | 2 | 3760 | 3416 | 352 | 25 | 23 | 3 | 9 | 8 | 1 |
| 2 | 9 | 12312 | 10264 | 2176 | 85 | 71 | 16 | 29 | 24 | 6 |
| 3 | 27 | 38936 | 34840 | 3534 | 267 | 241 | 25 | 90 | 81 | 9 |

Figure 2 a, b and c represents comparison between PCM, ULAW, AMR formats. Graphs shows that PCM and ULAW have little deviation with each other where ARM which encompasses compression inherently has small number of characters and SMS than other formats. In this case, all users spoke the sentence "Hello World" only one time. The average duration is about 1-2 seconds for this voice message.

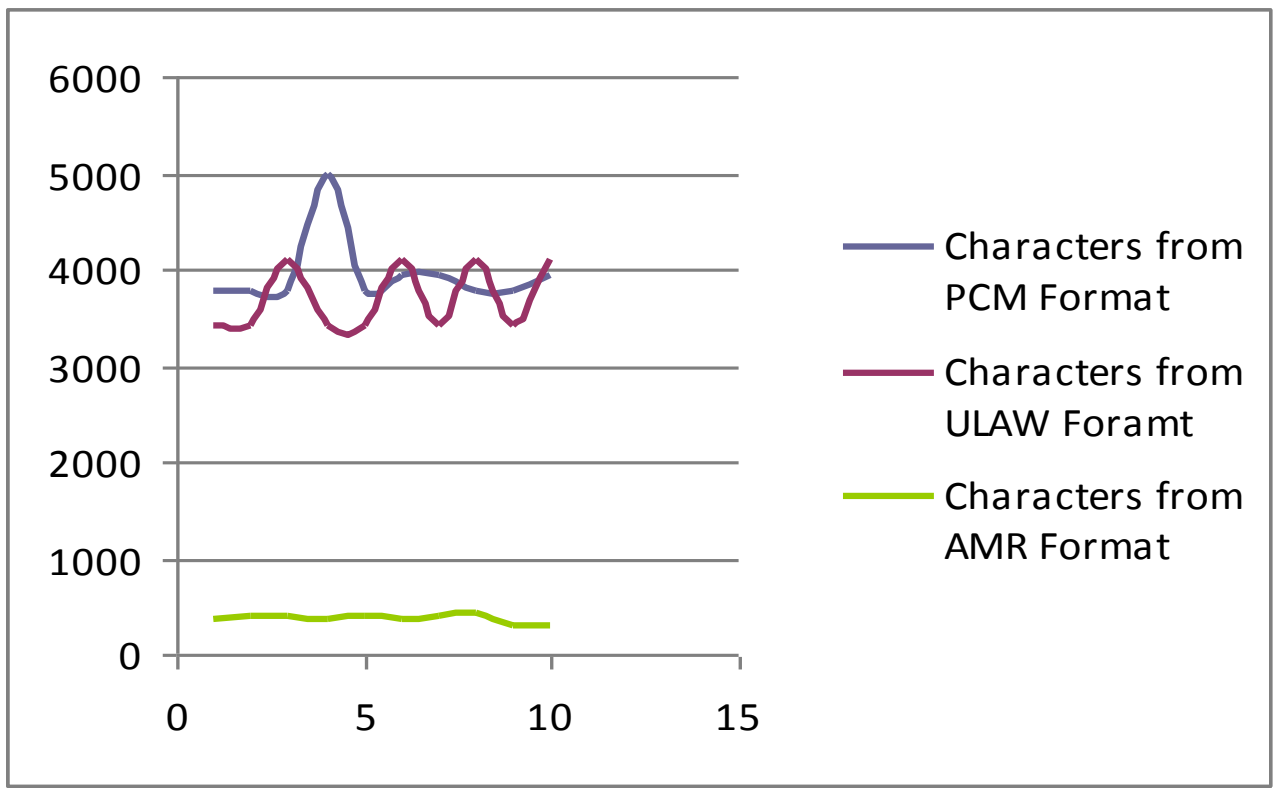

Fig 2 "a" Number of Extended ASCII characters for "*hello word*"

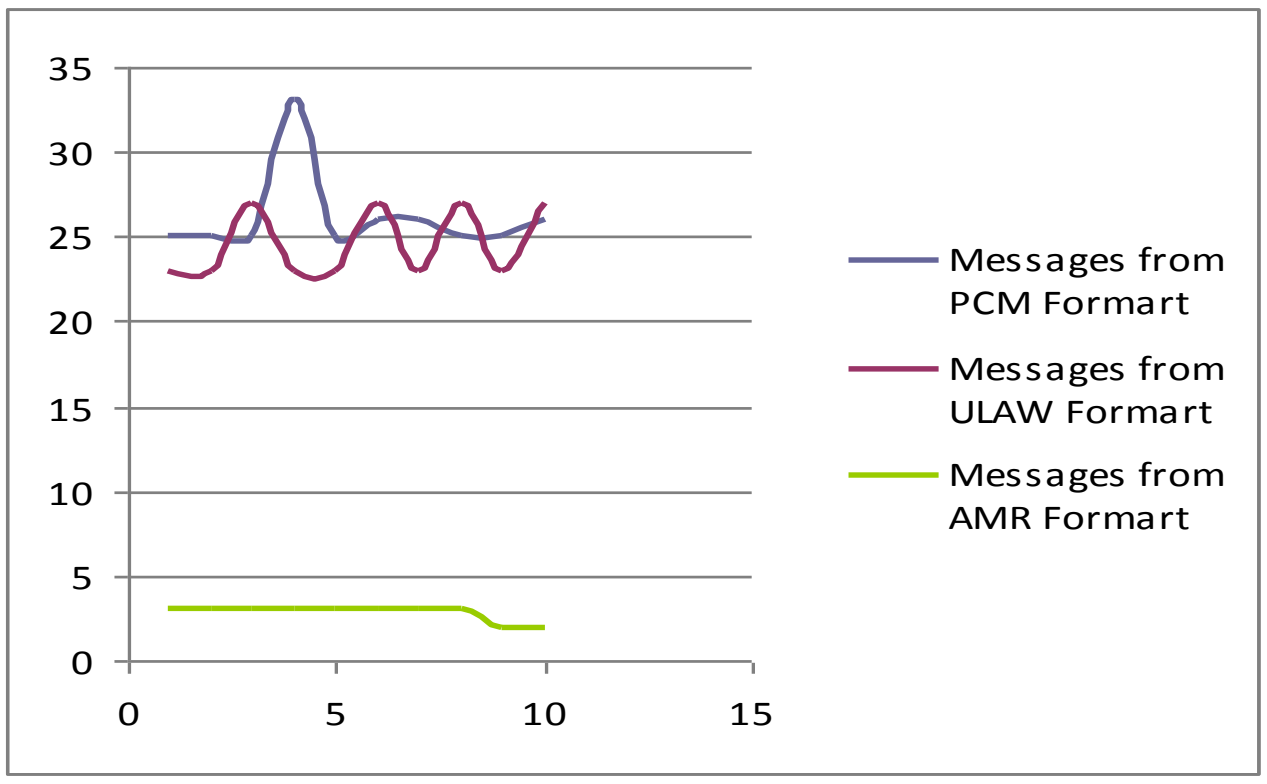

Fig 2 "b" Number of messages required for "*hello word*"

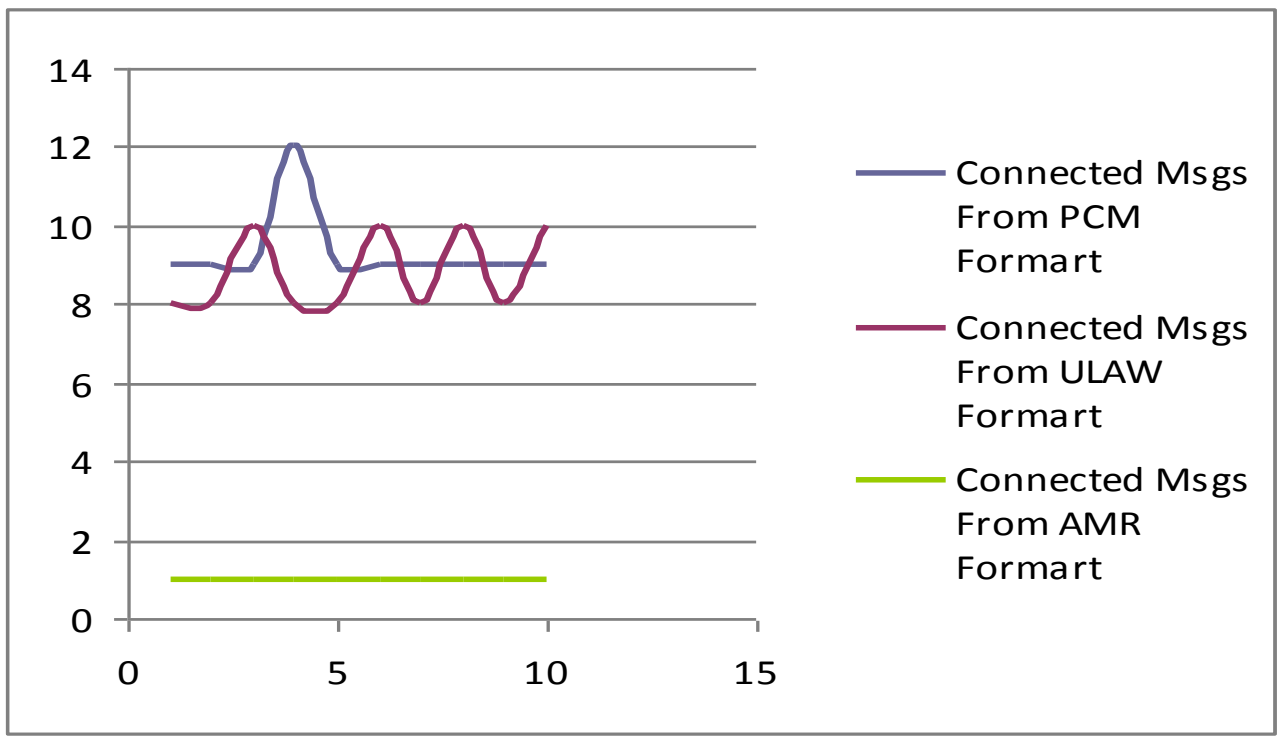

Fig 2 "c" Number of connected messages required for "*hello word*"

Increase number of spoken characters also increased the number of characters, messages and connected messages as well. In real world, SMS may lose or damage by any means can create problem for ordering SMS in order to play voice. For such scenario, our application plays voice messages on the basis of received messages only (loosely indexing is used).

Figure 3 a, b, c describe the impact of spoken words on the number of characters, messages and connected messages. In this experiment all user spoke the sentence "The quick brown fox jumps over the lazy dog" three times. The maximum duration is about 6 seconds for this message.

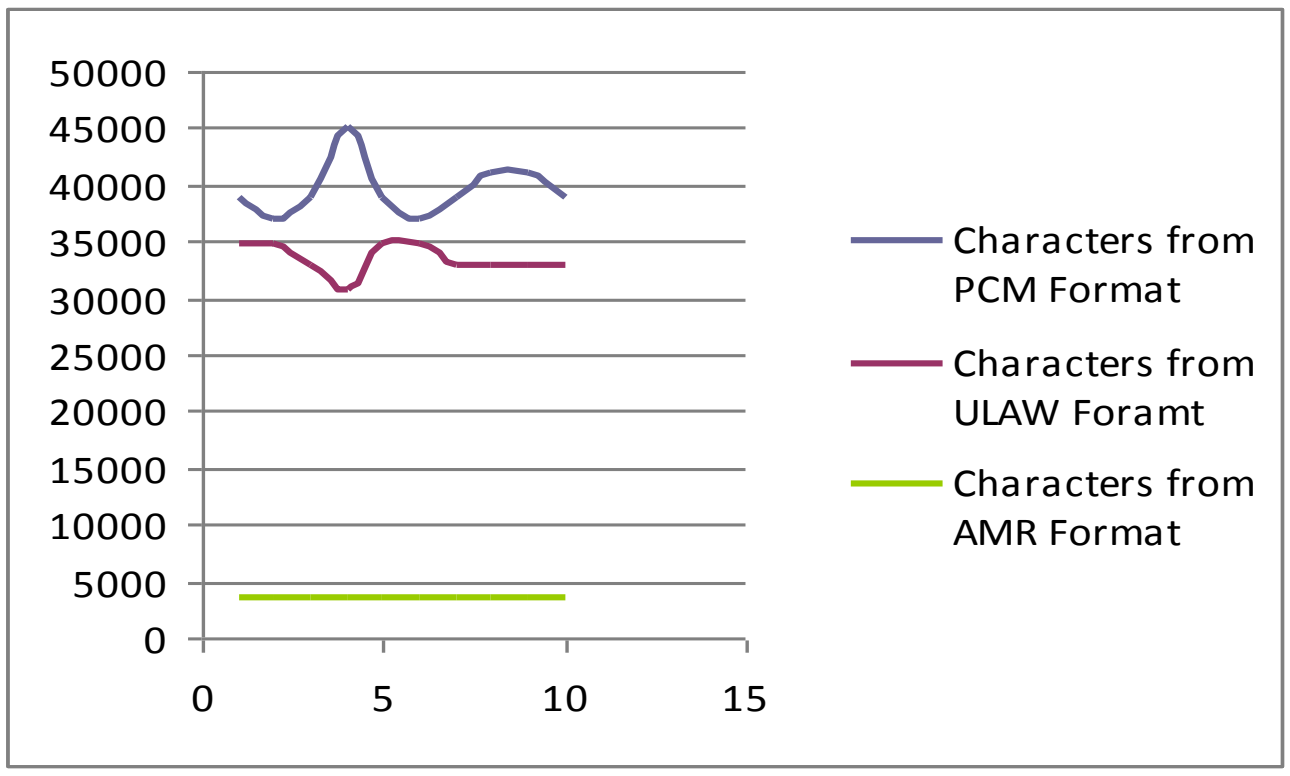

Fig 3 "a" Number of Extended ASCII characters for "*The quick brown fox jumps over the lazy dog*"

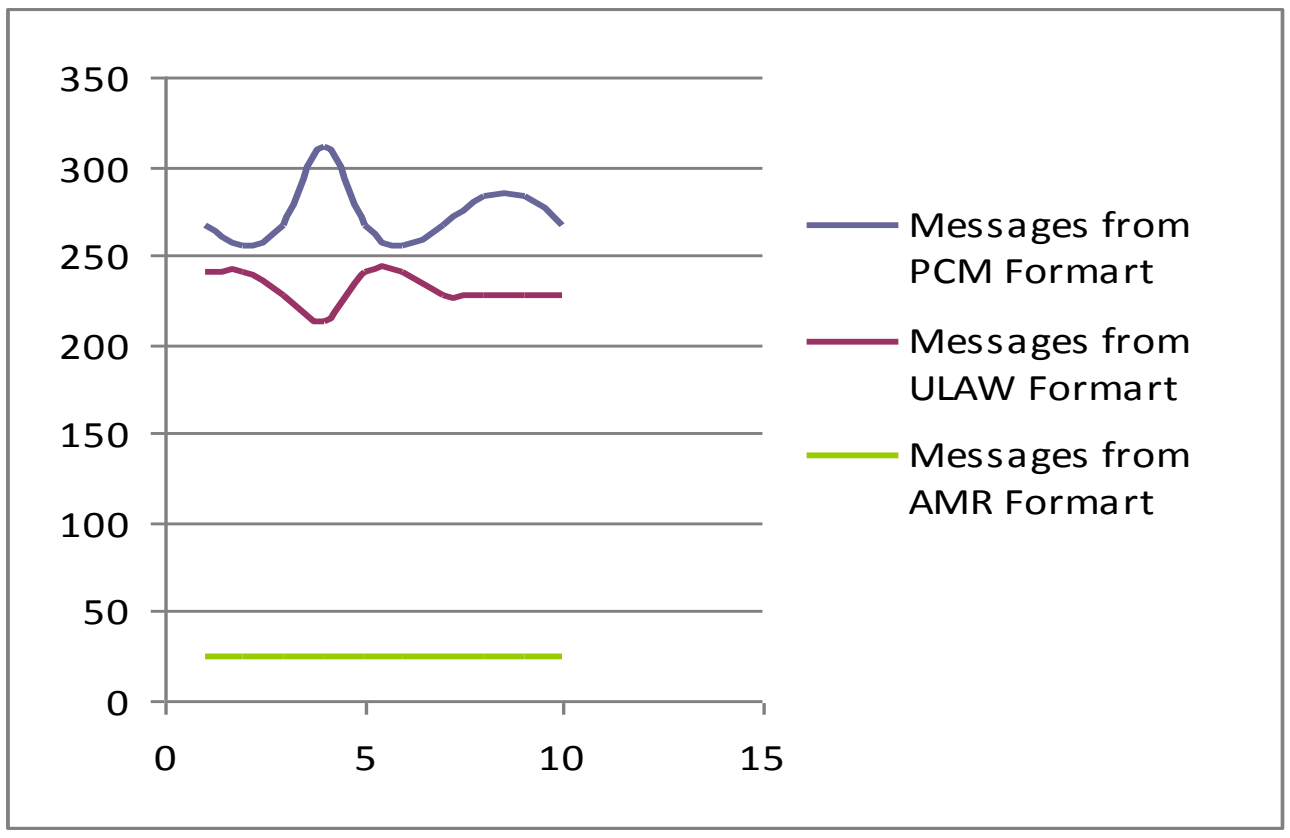

Fig 3 "b" Increase in number of messages for "*The quick brown fox jumps over the lazy dog*"

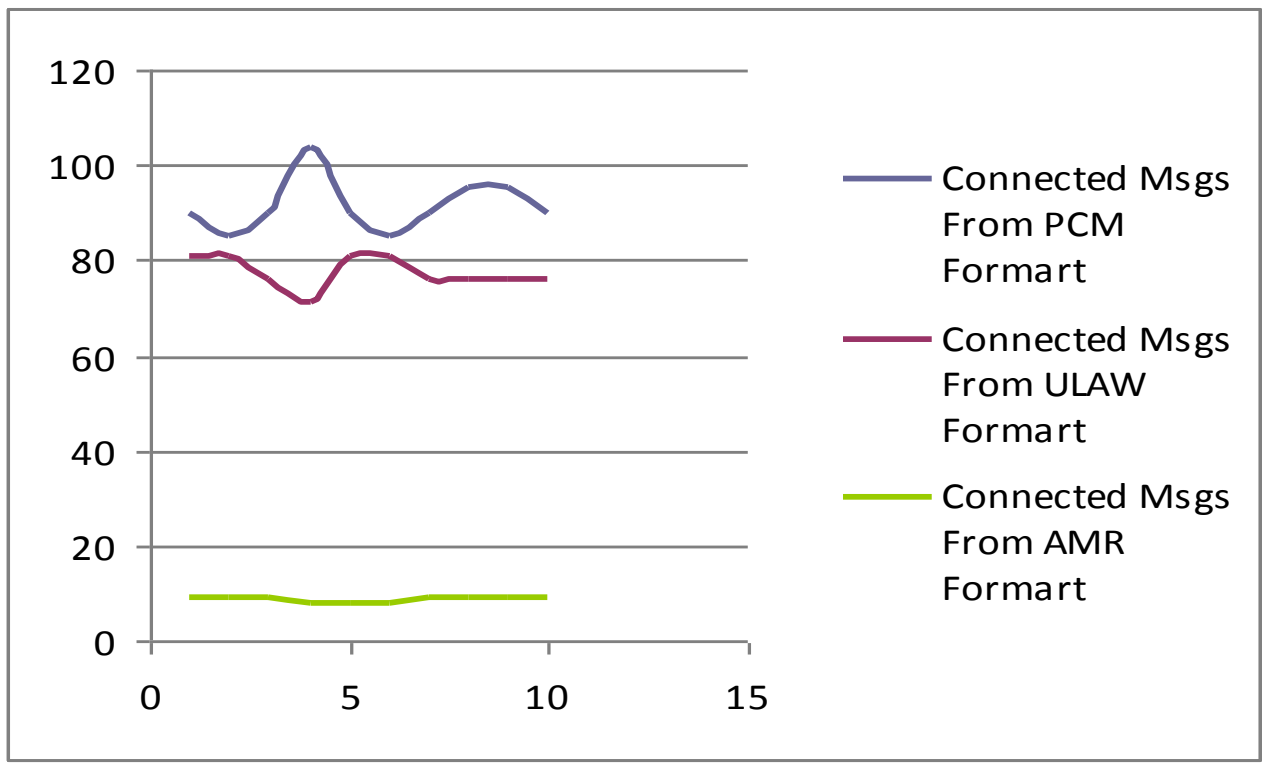

Fig 3 "c" Increase in number of connected messages for "*The quick brown fox jumps over the lazy dog*"

## 4 Conclusion

SMS is usually text based rather than MMS which contains text, voice, image and video etc. This paper presents a methodology of transmitting voice using SMS over GSM network. Using this application all GSM devices can send/receive voice messages without having capability of GPRS or 3G network. ARM format results show that usage of compression techniques may reduce number of messages and connected messages effectively which is a major drawback of the current proposed method.

## Future Work

In future we will apply different compression algorithms on the proposed methodology, compare them and lastly we will choose best compression algorithm for the said method. We will also apply this methodology for images, video etc.

**MR. Muhammad Fahad Khan** is working as a Lecturer and In charge of Management Information System in Federal Urdu University of Arts, Science and Technology, Islamabad Pakistan. He has number of journal publications. He is reviewer of two international journals .He is interested in Handheld Application Development, Information System Development, Network Security, Multimedia Communication and other their related fields. He did his Bachelor Degree from Federal Urdu University of Arts, Science and Technology in 2008 and now pursuing his Master Degree from IQRA University, Islamabad Campus.

**MS. Saira Beg** is working as a Research Associate (May 2009- up to date) in COMSATS Institute of Information Technology, Islamabad. Her Interest areas are Networks, Network Security and Artificial Intelligence and other their related Fields. She is the member of Artificial Intelligence Group at CIIT, Islamabad. She did her Bachelor Degree (Gold medalist) from Federal Urdu University of Arts, Science and Technology, Islamabad, Pakistan in 2008 and now doing her Master Degree from COMSATS Institute of Information Technology, Islamabad.